\documentclass[a4paper]{article}
\usepackage[english]{babel}
\usepackage[utf8]{inputenc}
\usepackage{amsmath}
\usepackage{amsthm}
\usepackage{amssymb}
\usepackage{latexsym}
\usepackage{tikz}
\usepackage[margin=1.03125in]{geometry}
\usepackage{graphicx}

\DeclareMathOperator*{\combine}{\bigcirc}

\title{Toward an enumeration of unlabeled trees}
\author{Recuero, P. M.}
\date{}

\begin{document}

\maketitle{}
\begin{abstract}
We present an algorithm that, on input $n$, lists every unlabeled tree of order $n$.
\end{abstract}
\section{Introduction}
The enumeration of structures is a primal theme of combinatorics. Whereas the problem of enumerating the labeled tree graphs
has been solved by Borchardt \cite{borchardt}---and subsequently improved by Cayley \cite{cayley}---over one century ago, an analogue for the unlabeled case
remained elusive.
Here, we present an algorithm that, on input $n$, lists every [unlabeled] tree of order $n$ exactly once.
\medskip\\
Before we begin, a couple remarks on notational conventions and terminology: a \textit{multiset} is a duple $S=(S^{*},M_{S})$,
where $S^{*}$ is a set, called the \textit{underlying set} and $M_{S}:S^{*}\to\mathbb{Z}^{+}$ is the so-called $multiplicity function$.
A partition $p$ of a natural number $n$ is a multiset $p=(p^{*},M_{p})$, $p^{*}\subset\mathbb{Z}^{+}$ such that $\sum_{i\in p^{*}} M_{p}i=n$.
We denote by $P(n)$ the set of all partitions of $n$, and exceptionally, cortiously let ({0},(0,1)) be the unique partition of zero.\\
Additionally, by $L_{n}$ we denote the linear [sometimes called ``path"] graph on $n$ vertices, that connected graph which either contains exactly two
vertices of degree one, all other vertices having degree two, or is the unique graph on one vertex.

\section{Main Result}
Our algorithm works in a series of steps, illustrated below for the case where our input $n=5$.

\begin{center}
\includegraphics[scale=0.25]{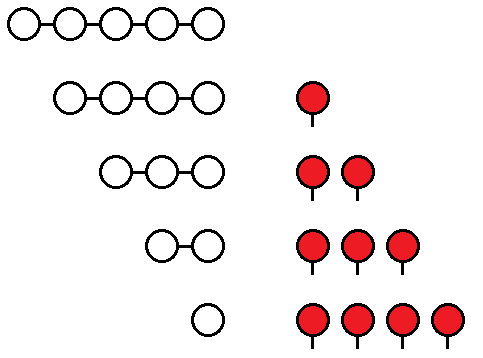}\\
\textit{Fig. 1}
\end{center}

That is, at the $k^{\text{th}}$ step, we have the linear graph $L_{n-k}$, which we call our \textit{backbone}, plus $k$ ``free" vertices.
Within each of these steps, we \textit{append} the $k$ free vertices to the backbone---that is, add an edge between a free vertex and
a vertex of the backbone---in every possible way, while avoiding repetitions and conserving diameter---\textit{i.e.} all trees generated
at step $k$ must have diameter $n-k$---, and then print out the resulting trees.

\begin{center}
\includegraphics[scale=0.25]{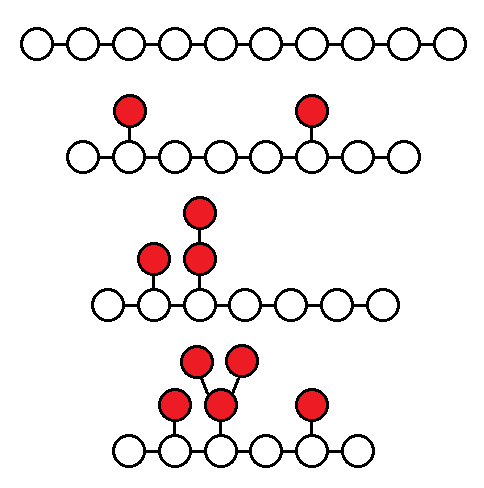}\\
\textit{Fig. 2}
\end{center}

Above we have a few examples of trees we might run into upon initialization of our algorithm on input $n=10$ at steps
zero, and three through five, respectively. Moreso, the figure above evinces a phenomenon to keep in mind: not only can we
append our free vertices \textit{directly} to our backbone, as in the second tree above, but we can also combine them with one another
before doing so, as in the third and fourth trees, forming \textit{compound appendices}; to the free vertices that were added directly
to our backbone, as in the second tree, we by contrast refer as \textit{singular appendices}. Appendices, compound or not, can also be
\textit{linear}, as in the second and third trees, or \textit{nonlinear}. Appendices that have already been connected to the backbone
we call \textit{substituents}.\\
But there is more than seemingly unnecessary terminological conventions to the figure above: as a matter of fact, instead of jumping straight into
our algorithm, we will visit the two simpler cases exemplified by the second and third trees above, viz. those where we can only append
the free vertices to our backbone directly or as halves of linear graphs, so as to warm us up and gather insight which shall later come to
us useful.

\subsection{Singular appendices}
How many ways are there to append $k$ free vertices directly to a backbone of order $n$? The naïve way tells us that there are $n\cdot k$
possibilities; however this of course leads to redundancies.

\begin{center}
\includegraphics[scale=0.25]{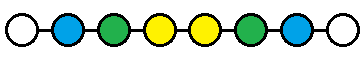}\\
\textit{Fig. 3}
\end{center}

In any given backbone, appending free vertices directly to vertices of same color as in the figure above results in the same tree. Of course,
if we only have one free vertex to append then we might as well just cut our backbone in half. Things do get more complicated as the number of
free vertices that are to be appended grows, however in fact, cutting our trees into ``half-trees" shall prove to be a didactic experiment, helping
us deal with the aforeseen symmetry.

\begin{center}
\includegraphics[scale=0.25]{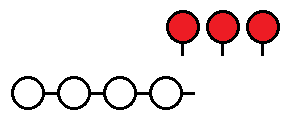}\\
\textit{Fig. 4}
\end{center}

Above we have a half-backbone of order four with three free vertices to be directly appended. Being devoid of symmetry, unlike their `complete'
congeners, the matter of appending a given number of free vertices directly to a half-backbone of a given order is trivial. Indeed, the figure
below hints at a mechanical method we can employ to generate all the possible trees that could arise therefrom, without repetitions.

\begin{center}
\includegraphics[scale=0.20]{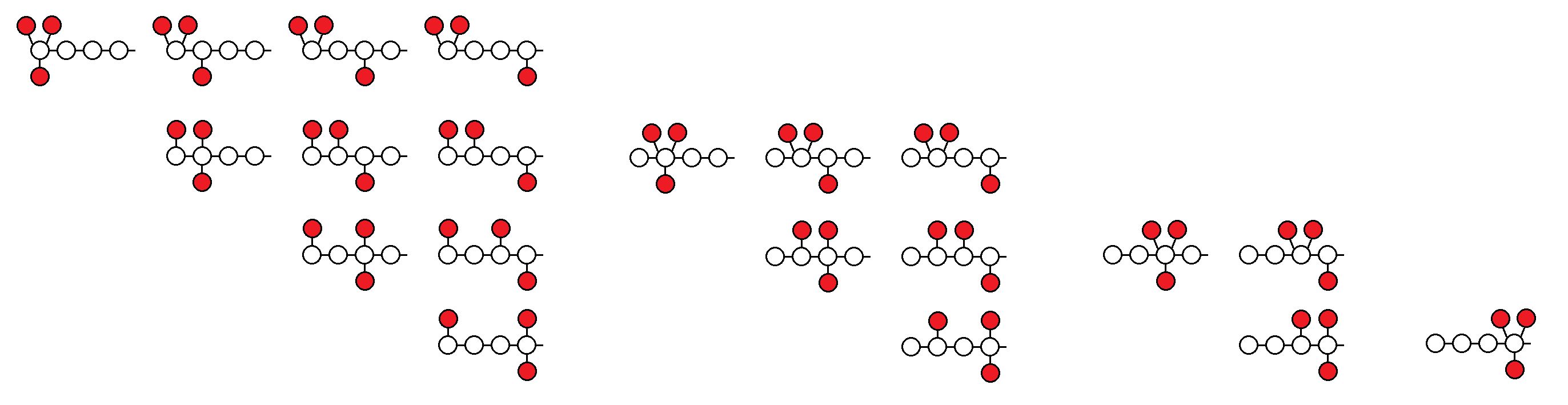}\\
\textit{Fig. 5}
\end{center}

And with some further thought, one can see that there are $G_{k}(n)$ ways to append $k$ free vertices directly to a half-backbone of order $n$, where

\begin{equation}
G_{0}(n)=
\begin{cases}
	0 & \text{if }n=0;\\
	1 & \text{if }n>0.
\end{cases}
\setlength{\parindent}{60pt}
\text{\indent}
\setlength{\parindent}{0pt}
G_{k+1}(n)=\sum_{i=0}^{n} G_{k}(i)
\end{equation}

Now to pass from half-backbones to full backbones, we proceed as follows: we regard a backbone of order $2n$ as two half-backbones, $h_{1},h_{2}$;
if we are to append $k$ free vertices directly to this backbone, then there are the following options: we can append $k$ free vertices directly to $h_1$
and zero to $h_2$; or we can append $k-1$ free vertices directly to $h_1$ and one to $h_2$; or $k-2$ to $h_1$ and two to $h_2$, and so on, up until
$k-\lfloor k/2 \rfloor$ to $h_1$ and $\lfloor k/2 \rfloor$ to $h_2$---if we went further than that then we would be going backwards---, as illustrated in the
figure below for the case where we have a backbone of order six, split into two half-backbones of order three, and four vertices to append.

\begin{center}
\includegraphics[scale=0.1875]{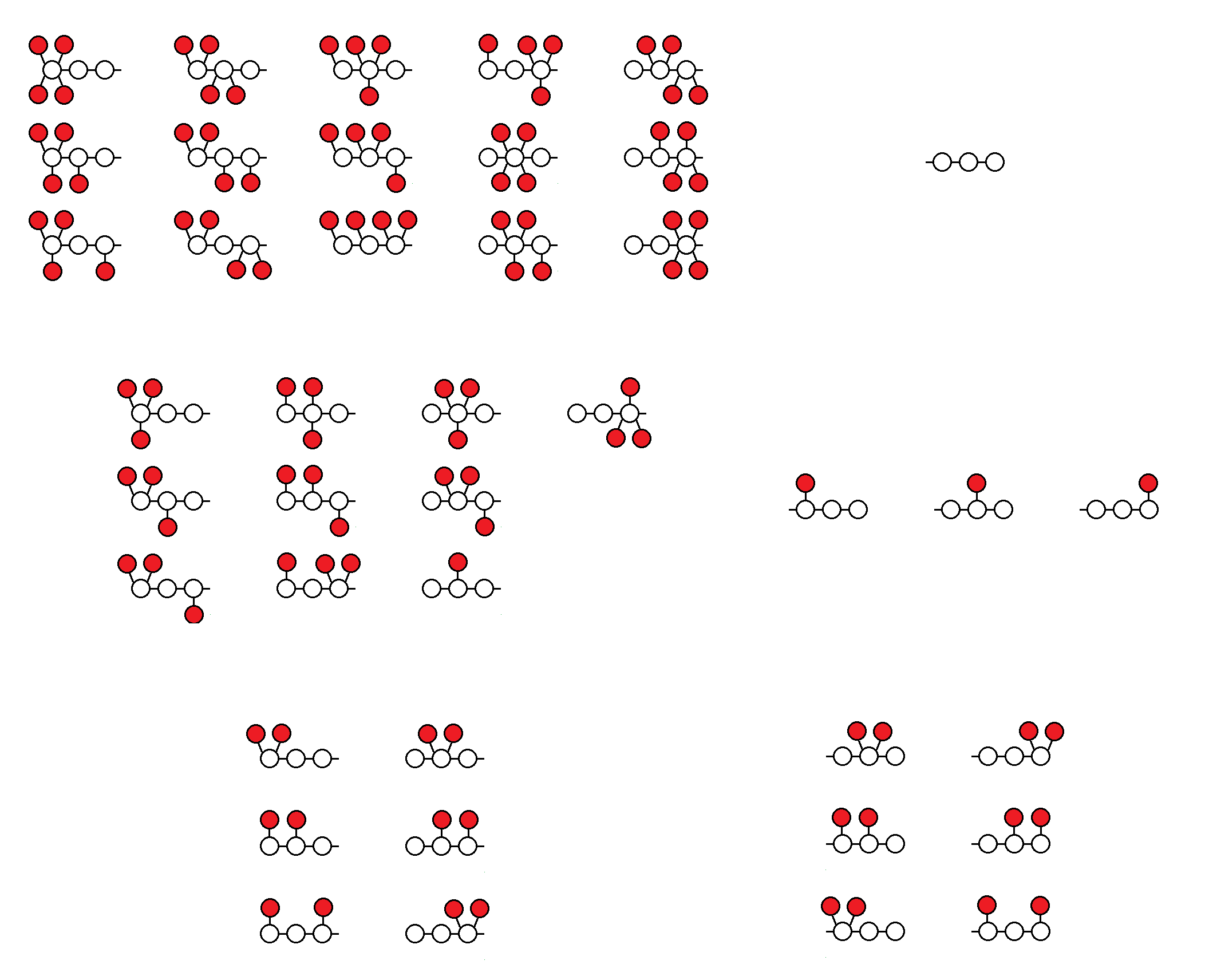}\\
\textit{Fig. 6}
\end{center}

And at each of these steps, we can ``glue" our half-trees into trees. But how many ways are there to do that? If we append $k_1$ free vertices to
$h_1$ and $k_2$ free vertices to $h_2$, then there are only two options:
\medskip\\

\setlength{\parindent}{10pt}

\indent (a) $k_{1}\neq k_{2}$, in which case, the sets of half-trees generated [from appending $k_{1}$ free vertices to $h_{1}$ and $k_{2}$ free
vertices to $h_{2}$] have no elements in common, and thus there are in total $G_{k_{1}}(n) \cdot G_{k_{2}}(n)$ trees that result from the gluing
of the half-trees---$n$ assumed to be the order of the half-trees;\\
\indent (b) $k_{1}=k_{2}$, in which case the sets are the same, and as such there are $\binom{G_{k_{1}}(n)}{2}+G_{k_{1}}(n)$ trees resulting
from the gluing of the half-trees.
\medskip\\
\setlength{\parindent}{0pt}

From the above we can thus derive that there are

\begin{equation}
\sum_{i=0}^{\lfloor k/2 \rfloor} F(k-i,i,n)
\label{n-fit-k-even}
\end{equation}

ways to directly append $k$ free vertices to a backbone of order $2n$, where

\begin{equation}
F(x,y,z)=
\begin{cases}
	G_{x}(z)\cdot G_{y}(z) & \text{if }x\neq y;\\
	\binom{G_{x}(z)}{2}+G_{x}(z) & \text{if }x=y.
\end{cases}
\end{equation}

is our gluing function.
\medskip\\

This however only solves the case for even backbones. To deal with odd backbones, we proceed as follows: we regard a backbone of order $2n+1$ as
two half-backbones of order $n$---the ``sides"--- and one extra vertex---the ``middle". And now we do as earlier: if we are to append $k$ free vertices
directly to an odd backbone, we can append $k$ free vertices to the sides and zero to the middle; or we can append $k-1$ free vertices to the sides
and one to the middle, and so on, up until zero free vertices to the sides and $k$ to the middle.
\medskip\\

As such we obtain that there are

\begin{equation}
\sum_{j=0}^{k\cdot d(n)} \sum_{i=0}^{\lfloor\frac{k-j}{2}\rfloor} F(k-j-i,i,\lfloor n/2 \rfloor)
\end{equation}

ways to append $k$ free vertices to a backbone of order $n$, where $d$ is just a function that maps its input to one, if it is odd, and to zero otherwise.

\begin{center}
\includegraphics[scale=0.25]{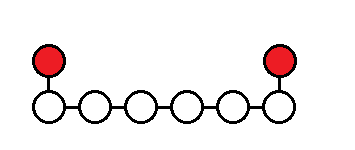}\\
\textit{Fig. 7}
\end{center}

There is yet another thing that we must note. Recall how we stated that our algorithm requires that all trees resulting from the appending of the free vertices
to the backbones have the same diameter as their backbones. Above we see a tree with diameter eight generated by the appending of two free vertices to a
backbone of diameter six. Were we to allow this genre of thing in our algorithm, we would have repetitions: indeed, were we to initialize our algorithm on input
$n=8$, then the tree above would be listead [at least] twice: once on the zeroth step, as the backbone of order eight with no substituents; and then again
at step two.\\
In our case dealing only with the direct appending of free vertices to backbones, the solution to this is simple: we just forbid substituents at end vertices
of our backbones. We shall see how things play out in other settings shortly.

\subsection{Linear appendices}
Here, we can combine our set of free vertices into halves of linear graphs before appending them to our backbone. First let us examine how the simplest cases
of this episode behave: those where all appendices are equal.\\
Notice how we can use the very same methods from the previous subsection here to append linear appendices to our backbones and half-backbones. The sole
difference lies in how we can commit to our conservation of diameter restriction.

\begin{center}
\includegraphics[scale=0.25]{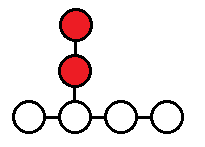}\\
\textit{Fig. 8}
\end{center}

The tree above was obtained from a backbone of diameter four, and yet displays diameter five, despite the fact that the substituent is not found on the end vertex.
Nonetheless, the solution to this is simple: we stipulate that appendices of radius $r$ must be appended to vertices of the backbone with distance equal to or greater than
$r$ from an end vertex, the end vertices themselves having distance zero.\\
With this, we can obtain that there are

\begin{equation}
\sum_{j=0}^{k\cdot d(n)} \sum_{i=0}^{\lfloor \frac{k-j}{2} \rfloor} F(k-j-i,i,\lfloor \frac{n-2r}{2} \rfloor)
\label{n-fit-k:r}
\end{equation}

ways to append $k$ appendices of radius $r$ to a backbone of order $n$ such that the resulting trees all have diameter $n$.
\medskip\\

However even though we are equipped to compute the number of and mechanically generate trees obtained by the appending of linear appendices to backbones,
we can only do so when all appendices in question are all of equal radii. Let us see how things go when we are without this prerogative.\\
The first thing to note is that, evidently, the dividing of a set of $k$ free vertices into halves of linear graphs coincides with integer partitions of $k$.\\
Now suppose we are faced with the following:

\begin{center}
\includegraphics[scale=0.25]{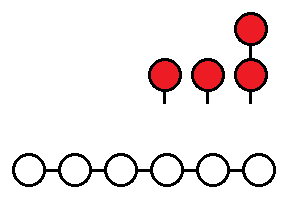}\\
\textit{Fig. 9}
\end{center}

We know how to both generate [while conserving diameters] the trees from the appending of the two singular vertices and from the appending of the one half line graph
of order two, though so far only separately.

\begin{center}
\includegraphics[scale=0.25]{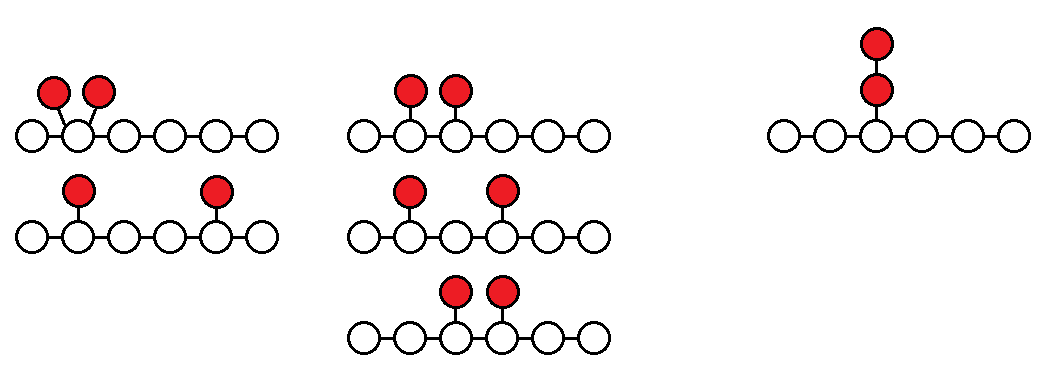}\\
\textit{Fig. 10}
\end{center}

To attain simultaneity, we can ``combine" the trees generated from appendices of different radii.

\begin{center}
\includegraphics[scale=0.25]{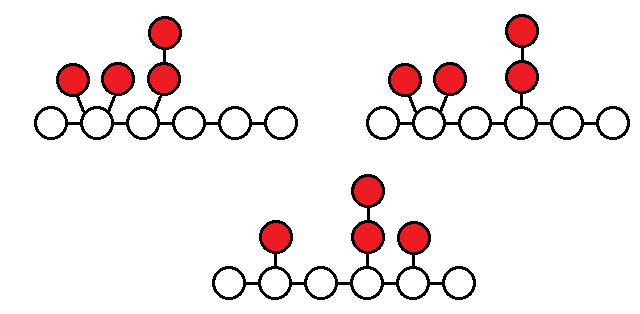}\\
\textit{Fig. 11}
\end{center}

Above we see the results of combining the top-right tree with the top-left and bottom trees [Fig. 10]. One can see that there are two ways to combine the top-right tree
with the top-left one, though only one way to combine the top-right one with the bottom one. Evidently this is a matter of symmetry. Without much thought one can notice that
whenever a symmetric tree is combined with another tree---be this second one symmetric or asymetric---there is only one possible combination, and when two asymmetric trees
are combined, there are two possibilites.\\
Let us then introduce the $\circ$ operation: for two sets $S,T$ of trees of same diameter, let $S\circ T$ denote the set of all trees resulting from combining $s\in S$ with $t\in T$;
patently we have that

\begin{equation}
|S\circ T| = 2\cdot\alpha(S)\cdot\alpha(T) + \sigma(S)\cdot\alpha(T) + \alpha(S)\cdot\sigma(T) + \sigma(S)\cdot\sigma(T)
\end{equation}

Where by $\sigma(S),\alpha(S)$ we denote the number of symmetric, resp. asymmetric trees in $S$.\\
Let us denote by $H(k,r,n)$ the set of all trees generated by appending $k$ linear appendices of radius $r$ to a backbone of order $n$ without altering its diameter;
we can calculate this set's cardinality with equation \eqref{n-fit-k:r}; however if we are interested in the cardinality of the set of trees generated by combining several
$H(k_{i},r_{i},n)$'s, then we also need to calculate their $\alpha$ and $\sigma$. Clearly, if $k$ is odd, then $\sigma(H(k,r,n))=0$, and if $k$ is even, then
$\sigma(H(k,r,n))=G_{k/2}(\lfloor n/2 \rfloor)$; from these, $\alpha(H(k,r,n))$ can trivially be derived.\\
Armed with these, we can calculate the number of trees that arise from the combination of two sets of trees generated by the appending of halves of linear graphs of two different
radii; however we will also want to calculate combinations of combinations of several sets of graphs. By noting that the combination of an asymmetric tree with any other tree---be it
symmetric or asymmetric---yields an asymmetric tree, and that the combination of two symmetric trees yields a symmetric tree, we can determine that

\begin{equation}
\sigma(|S\circ T|) = \sigma(S)\cdot\sigma(T)
\end{equation}

whence again $\alpha(|S\circ T|)$ can trivially be inferred.
\medskip\\

There is just one last thing with which we must deal before concluding our work here, and as in the previous subsection, it pertains to odd graphs and their middles. In the case where
all appendices were singular, we proceeded by gradually increasing the number of free vertices directly appended to the middle; but in this here case, we are dealing with partitions,
and if we just increase those blindly, mistakes can happen.

\begin{center}
\includegraphics[scale=0.25]{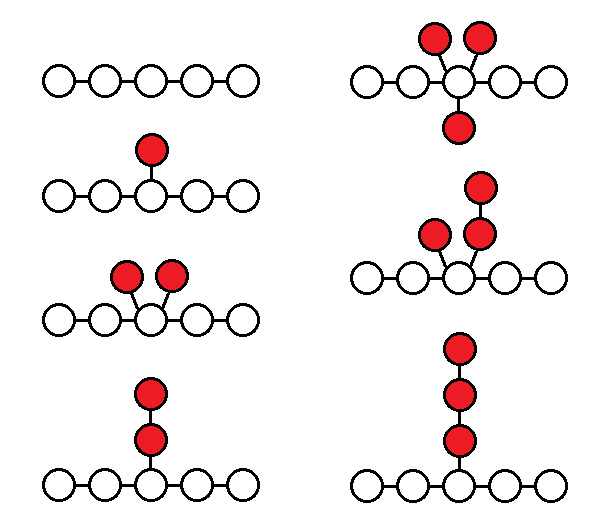}\\
\textit{Fig. 12}
\end{center}

In the bottom-left tree above, for example, the middle substituent has radius greater than the distance between the vertex to which it is appended and an end vertex, violating our rule;
indeed, even though it was generated from a backbone of diameter five, it displays diameter six. We need, as such, to discard those partitions.\\
Putting all of the above together, we can obtain that there are

\begin{equation}
\sum_{j=0}^{k\cdot d(n)} \sum_{J\in P(j)} \Bigg[ C(J^{*},n) \cdot \sum_{Q\in P(k-j)} \Big| \combine_{q\in Q^{*}} H(M_{Q}(q),q,n) \Big| \Bigg]
\end{equation}

trees generated by appending a set of $k$ free vertices as halves of linear graphs to a backbone of order $n$, where

\begin{equation}
C(S,n)=
\begin{cases}
	1 & \text{if }\max_{i\in S} \leq \lfloor n\rfloor;\\
	0 & \text{if }\max_{i\in S} >    \lfloor n\rfloor.
\end{cases}
\end{equation}

is the function that prevents appendices that are ``too long" to be appended to the middle vertex by checking the greatest element of the underlying set
of the given partition---corresponding to the appendix of greatest radius. $\Box$


\subsection{Nonlinear appendices}
We are  now ready to tackle the case where the free vertices are allowed to combine with one another in any way they please. The first challenge thus to us presented is finding
all possible such combinations---of compound appendices. Note how compound appendices coincide with half-trees. The following is an overview of a procedure to list all half-trees
of a given order.

\begin{center}
\includegraphics[scale=0.25]{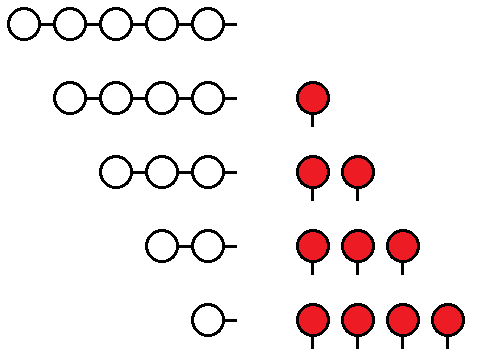}\\
\textit{Fig. 13}
\end{center}

Just as we said that our algorithm for the listing of trees would be divided in steps, so will be our algorithm for the listing of half-trees: at the $k^{\text{th}}$ step we have
a (half-)backbone of order $n-k$ and $k$ free vertices, $n$ being our input; illustrated above is the case for $n=5$. Within each of those steps we do as follows: we go over the
partitions of $k$, and for each of those we ``apply" them to our set of $k$ free vertices. That is, if $k=15$, say, and we have $2+2+2+3+3+3$ as our given partition, then
we shall divide our $k$ free vertices into three half-trees of order two, and three half-trees of order three.

\begin{center}
\includegraphics[scale=0.25]{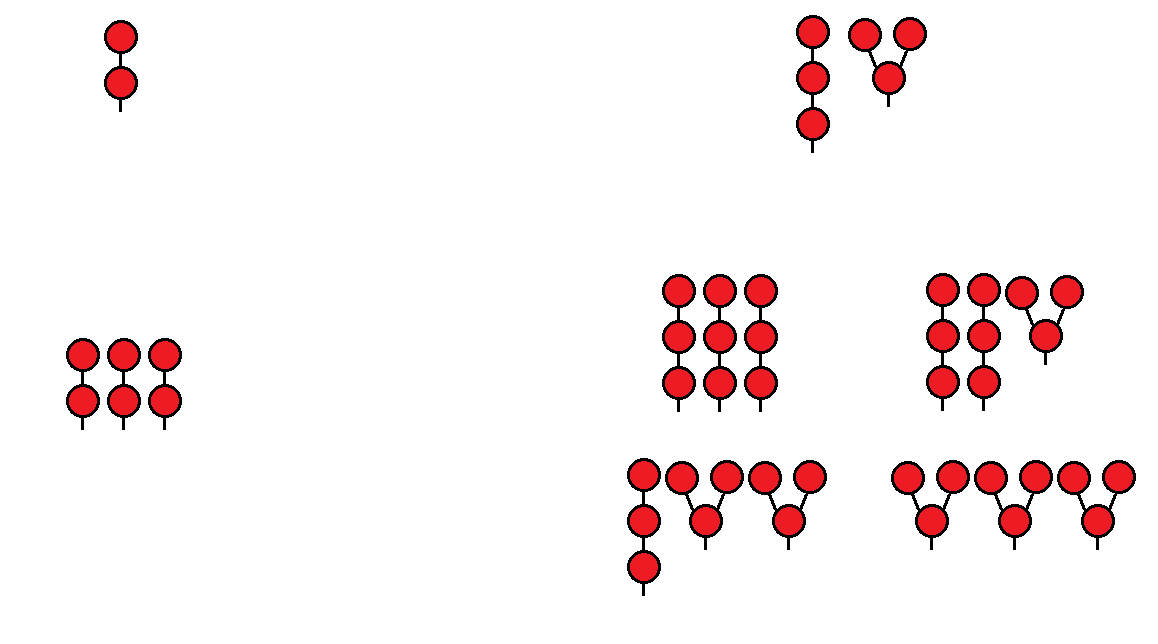}\\
\textit{Fig. 14}
\end{center}

And for each of those sets of half-trees of same order, we can choose from the set of all half-trees of that given order to form our \textit{fixed order half-three multiset},
as seen above for the case of the aforementioned partition on $k=15$: there are only two half-trees of order three; in our partition we must have three of those, so naturally
there are $\left(\!\!\binom{2}{3}\!\!\right)=4$ possible fixed order half-three multisets to choose from. Meanwhile, there is only one half-three of order two. Now, we can choose
any combination of fixed order half-tree multisets to ``manifest" our partition, yielding us our \textit{appendix set}, itself a multiset whose underlying set has half-trees as elements,
the union of its fixed order half-tree multiset constituents.\\

\begin{center}
\includegraphics[scale=0.25]{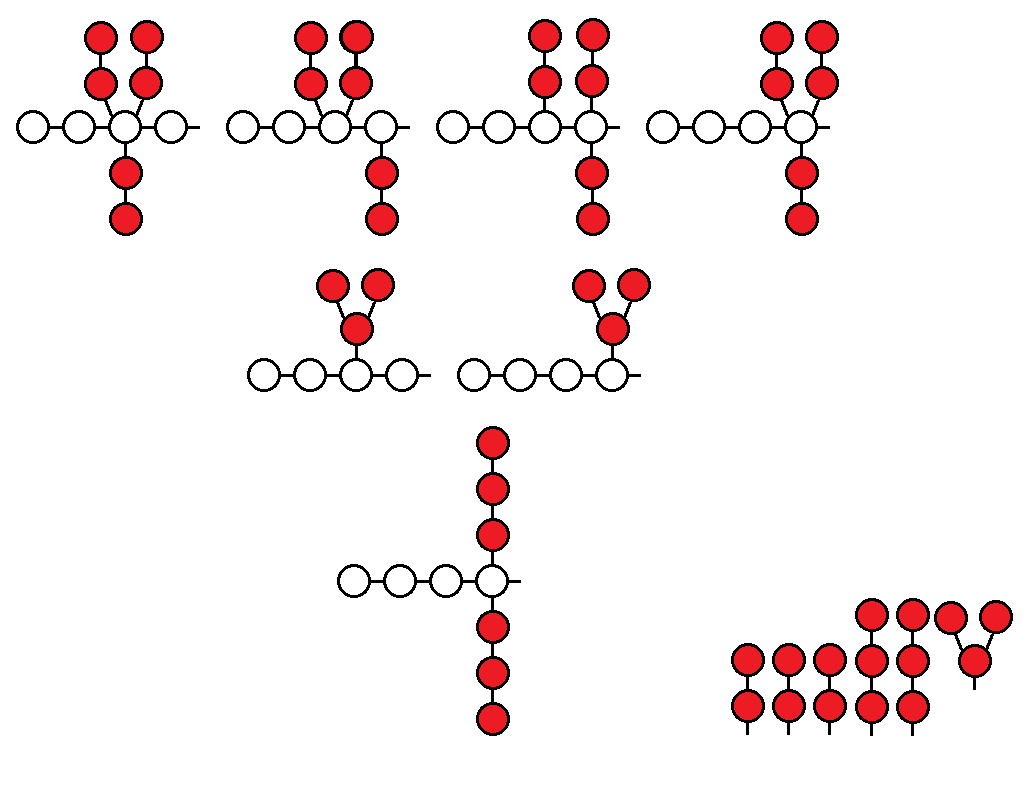}\\
\textit{Fig. 15}
\end{center}

We go over all appendix sets, and for each $A$ of them, for each $a\in A^{*}$, we list all of the half-trees generated by appending $M_{A}(a)$ appendices $a$ to our half-backbone in question
while conserving radius. In the figure above, for example we have chosen one of the appendix sets from Fig. 14 and apply it to a half-backbone of order four.\\
So far, everything is just like earlier when all substituents were linear. Indeed, most of the times we can treat nonlinear substiuents in the same manner we treat linear ones, but as we shall
see later they have a behavior of their own. For the purpose of this overview, we can simply ignore that and assume that we already know how to deal with them.\\
Now, as earlier, we combine our half-trees and obtain our set of half-trees generated from appendix set $A$ applied to the given half-backbone.
\medskip\\
Recapitulating in a clumsy mixture of prose and pseudocode,
\begin{verbatim}
HT_n := { }
for k in [0,n-2]:
    for p in P(k):
        for AS generatable from p:
            for a in AS:
            generate h.-t.'s by appending M_AS(a) a's to half-backbone of order n-k
        combine h.-t.'s generated from different a's
        append the results to HT_n
\end{verbatim}

Where \verb+HT_n+ is meant to be the set of all half-trees of order $n$, to which we add the sets of half-trees generated by appending $k$ free vertices,
combined as per every appendix set generatable from every partition of $k$ to half-backbones of order $n-k$. $\Box$
\medskip\\

And now we turn to the aforementioned peculiarity of nonlinear appendices.

\begin{center}
\includegraphics[scale=0.25]{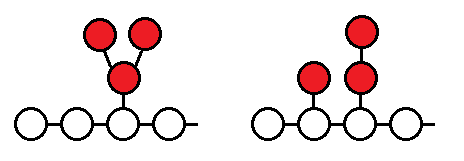}\\
\textit{Fig. 16}
\end{center}

Consider the two half-trees above. Each was generated by appending three free vertices to a half-backbone of order four, though by different appendix sets and partitions,
and as such---were we to ignore the peculiarities of nonlinear appendices; above we have assumed that we already know how to deal with those---would both be listed,
were we to initialize our algorithm for listing half-trees on input $n=7$, at the third step. However note how they are isomorphic.\\
Let us call a substituent of radius $r$ that is appended to a vertex of the half-backbone with distance exactly $r$ from the half-backbone [such as the nonlinear substituent on the tree
to the left in the figure above] \textit{terminal}. One can easily verify that often times, when appending nonlinear substituents terminally, the resulting half-tree is isomorphic to another
half-tree generated from the same number $k$ of free vertices, but with different appendix set. Other times, this does not happen, as in the top tree in the figure below.

\begin{center}
\includegraphics[scale=0.25]{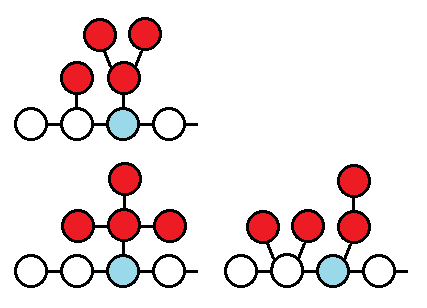}\\
\textit{Fig. 17}
\end{center}

To deal with this, we will introduce an ordering relation $<_{ht}$ on the half-trees. In a tree with a terminal substiuent, call the half-tree formed by cutting the edge between the vertex
of the half-backbone to which the substituent in question is appended and its neighbor closer to the end vertex---in Fig. 17 above, merely everything to the left of the blue vertex---of the half-backbone the \textit{tip} induced by that substituent. We stipulate that an appendix $a$ can only be appended terminally if we have that $a\leq_{ht}t$, where $t$ is the tip it shall thereby induce.
\medskip\\
The idea behind $<_{ht}$ shall be simple. Consider an expanded version of the aforementioned algorithm for the listing of all half-trees of a given order wherein all half-trees of all orders are listed:
we simply go over the positive integers, starting from one, and run the algorithm above at each of those. We will say that for two half-trees $a,b$, we have that $a<_{ht}b$ if by this expanded algorithm, $a$
is listed before $b$.\\
As such it is required of us that we order the \verb+for+'s within our first algorithm, define an ordering on the partitions of a given integer, and then on the appendix sets of a given partition. To order
these last ones in turn, we need an ordering on the fixed order half-tree multisets.\\
Let us begin with the ordering $<_{p}$ on the partitions. To decide the relationship between two given partitions $p,q$ of the same integer, we proceed as follows:

\begin{verbatim}
1.  X_p := p*
2.  X_q := q*
3.  if max{i in X_p} < max{j in X_q}:
        then p < q
    else:
        if M_p(max{i in X_p}) < M_q(max{j in X_q}:
            then p < q
        else:
            X_p -= max{i in X_p}
            X_q -= max{j in X_q}
            go to step 3
\end{verbatim}

That is, we compare the greatest part of the partitions, and then its multiplicity, and then the second greatest part, and then its multiplicity and so on. In the procedure described above,
at some point, \verb+X_p+ and \verb+X_q+ (our containers) will become empty, and then, seeing as all previous comparison of parts and their multiplicities led to a tie, the partitions will be equal. It also doesn't hurt to notice that it is impossible for one of the \verb+X+'s
become empty before the other---if a partition $p$ has fewer parts than another partition $q$ on the same integer, by our definition we will have $p>_{p}q$ and our algorithm will decide
that before we run out of parts.
\medskip\\
Next, an ordering on the appendix sets. First, only appendix sets generated from the same partition are important to us. This ordering here will be similar to our ordering on the partitions:
recall that an appendix set consists of fixed order half-three multisets, each derived from a part of the partition whence the appendix set was generated. Assuming an ordering on these
last ones, we simply proceed just like we did for partitions: for two appendix sets $A,B$ generated from the same partition $p$, we compare first the fixed order half-three multiset derived
from $\max_{i\in p*} i$ and so on.
\medskip\\
Thus, on to the ordering $<_{m}$ on the fixed order half-three multisets. Let us denote by $HT_n$ the set of all half-trees of order $n$. To each $h\in HT_n$ we assign a prime number as follows:
to the first half-tree that is listed by the first half-tree listing algorithm (the one that lists half-trees of a given input), $2$ is assigned; to the second, $3$ is assigned, and so on. For a half-tree
$h$, let us denote by $\nu(h)$ the prime number thereto assigned. Now for a fixed order half-tree multiset $H$ we let

\begin{equation}
N(H)=\prod_{h\in H^{*}}{\nu(h)}^{M_{H}(h)}
\end{equation}

Now we define that for two fixed order half-tree multisets (of half-three of same order) $H_{1},H_{2}$, we have that $H_{1}<_{m}H_{2}$ iff $N(H_{1})<N(H_{2})$.$\Box$
\medskip\\

Now, at first sight it might seem self-referential, impossible or illegal devise an ordering on the half-trees based on an ordering on fixed order half-tree multisets that is itself based on the ordering
on the half-trees. However notice that the smallest half-trees, by virtue of being simple, are greater or lesser than one another not because of the fixed order half-tree multisets that make up
their appendix set, but rather because of petty differences in their radii or even partitions on their free vertex set: these minimal trees are too little to make room for such complexity.
As our half-trees grow and the first half-trees start to exhibit big free vertex sets allowing diverse appendix sets, these last ones, in their turn, will be made out of the simpler half-trees, which
will have already been ordered, and induce themselves an ordering on the larger half-trees.
\medskip\\

Armed with all of the above, we are now able to go in greater depth into our half-tree listing algorithm, whence we shall subsequently, at last, derive our tree listing algorithm. Having
fixed $n$ our input,
\medskip\\
\begin{samepage}
$\bullet$ For natural $k\in [0,n-2]$:\\
\setlength{\parindent}{15pt}
\indent$\bullet$ For each $K\in P(k)$:\\
\indent\indent$\bullet$ For each appendix set $A$ generatable from $K$:\\
\indent\indent\indent$\bullet$ For each $a\in A^{*}$:\\
\indent\indent\indent\indent$\bullet$ Generate $H'(M_{A}(a),a,n-k)$;\\
\indent\indent\indent$\bullet$ Combine all of the $H'(M_{A}(a),a,n-k)$'s;\\
\indent\indent\indent$\bullet$ List the half-trees in the resulting set.\\
\setlength{\parindent}{0pt}
\end{samepage}

A few comments are warranted:\\
--- In keeping with the orderings we devised above, at the second, third and fourth bullets we go over the partitions, appendix sets and half-trees respectively in increasing order;\\
--- Here, $H'$ is an extension of our $H$ from the previous subsection; whereas the latter only took numbers as inputs---seeing as all graphs with which it dealt coincided with numbers---the former's second argument is a half-tree; as such, $H'(k,a,n)$ is the set of all half-trees generated by appending $k$ half-trees $a$ to a half-backbone of order $n$ while conserving radius (here we allow nonlinear appendices to be appended terminally, for our half-trees are not done yet);\\
--- It is in the combine step that lie the nottiness the nonlinear appendices bring to the table; trees without terminal nonlinear substituents can be combined freely,
however when combining a tree with terminal nonlinear substituents $a_{1},a_{2},\text{...}$ , it must be verified if the induced tip $t$ is such that $t>_{ht}a_{i}$. As such, it must be the case that when combining the $H'(k,a,n)$'s, we begin from the half-trees of substituents of least radius, so that when the time comes to deal with a half-tree with a nonlinear terminal substituent, all appendices that could have been appended to the tip have already been;\\
--- At last the listing of the half-trees must be orderly. For many half-trees arise from the same appendix set, partition
and half-backbone and yet we must order these as well. Here, we do so according to the position of the substituents. The simple idea is that the trees that have the smallest (by $<_{ht}$) substituents closer to the end are lesser, in keeping with the the mechanical method we employ to append a given number of free vertices onto a half-backbone (cf. Fig. 5 and many others). This is solved for the case were all substituents are equal, however we still must with half-trees with varied substituents. We do this as follows: for two trees with the same substituents $a_{1}<_{ht}a_{2}<_{ht}\text{...}<_{ht}a_{n}$, we first verify how close to the end vertex the $a_{1}$ substituents are in both trees; there being a tie there, we go on to consider the $a_{2}$ substituents and so on. $\Box$\pagebreak

At last, our algorithm for the listing of trees---which by now should look like an obvious extension of the one for half-trees above, seeing as the only additions are the steps where we fuse the half-trees and deal with the middle vertices:
having $n$ our input been fixed,
\medskip\\

$\bullet$ For natural $k\in[0,n-3]$:\\
\setlength{\parindent}{15pt}
\indent$\bullet$ If $n-k$ is even:\\
\indent\indent$\bullet$ For $K\in P(k)$:\\
\indent\indent\indent$\bullet$ For each appendix set $A$ generatable from $K$:\\
\indent\indent\indent\indent$\bullet$ For each $a\in A^{*}$:\\
\indent\indent\indent\indent\indent$\bullet$ For natural $i\in[0,\lfloor\frac{M_{A}(a)}{2}\rfloor]$:\\
\indent\indent\indent\indent\indent\indent$\bullet$ Generate $H'(M_{A}(a)-i,a,\frac{n-k}{2} )$;\\
\indent\indent\indent\indent\indent\indent$\bullet$ Generate $H'(i,a,\frac{n-k}{2})$;\\
\indent\indent\indent\indent\indent\indent$\bullet$ Fuse the $H'$'s into trees, yielding a set $FH_{a}$ of trees;\\
\indent\indent\indent\indent$\bullet$ Combine all of the $FH_{a}$'s;\\
\indent\indent\indent\indent$\bullet$ List the trees in the resulting set.\\
\indent$\bullet$ If $n-k$ is odd:\\
\indent\indent$\bullet$ For natural $j\in[0,k]$:\\
\indent\indent\indent$\bullet$ $FH_{j} := \emptyset$\\
\indent\indent\indent$\bullet$ For $Q\in P(k-j)$:\\
\indent\indent\indent\indent$\bullet$ $FH_{Q} := \emptyset$\\
\indent\indent\indent\indent$\bullet$ For each appendix set $B$ generatable from $Q$:\\
\indent\indent\indent\indent\indent$\bullet$ For each $b\in B^{*}$:\\
\indent\indent\indent\indent\indent\indent$\bullet$ For natural $i\in[0,\lfloor\frac{M_{B}(b)}{2}\rfloor]$:\\
\indent\indent\indent\indent\indent\indent\indent$\bullet$ Generate $H'(M_{B}(b)-i,b,\lfloor\frac{n-k}{2}\rfloor)$;\\
\indent\indent\indent\indent\indent\indent\indent$\bullet$ Generate $H'(i,b,\lfloor\frac{n-k}{2}\rfloor)$;\\
\indent\indent\indent\indent\indent\indent\indent$\bullet$ Fuse the $H'$'s into trees, yielding a set $FH_{b}$ of trees;\\
\indent\indent\indent\indent\indent$\bullet$ Combine all of the $FH_{b}$'s, call the resulting set of trees $FH_{B}$;\\
\indent\indent\indent\indent\indent$\bullet$ $FH_{Q} = FH_{Q}\cup FH_{B}$\\
\indent\indent\indent\indent$\bullet$ $FH_{j}= FH_{j}\cup FH_{Q}$\\
\indent\indent\indent$\bullet$ For every $J\in P(j)$:\\
\indent\indent\indent\indent$\bullet$ For every appendix set $C$ generatable from $J$:\\
\indent\indent\indent\indent\indent$\bullet$ Check if the $c\in C^{*}$ of greater radius has radius greater than $\lfloor n/2\rfloor$;\\
\indent\indent\indent\indent\indent$\bullet$ If it does, discard this appendix set and move on to the next; if not, proceed;\\
\indent\indent\indent\indent\indent$\bullet$ Append the appendices to the middle vertex of a backbone of order $n-k$ as per $C$;\\
\indent\indent\indent\indent\indent$\bullet$ Combine the resulting tree to all of those in $FH_{j}$;\\
\indent\indent\indent\indent\indent$\bullet$ List the resulting trees;
\setlength{\parindent}{0pt}
\begin{flushright}$\blacksquare$\end{flushright}
\section{Epilogue}
Evidently, from here, the elaboration of a closed-form expression for the number of trees listed by our algorithm on a given input, whence an enumeration on the unlabeled tree graphs would be yielded, presents an interesting task; such shall be the subject of a future writing. In passing, an implementation of the algorithm is also on its way.


\begin{thebibliography}{1}
\bibitem{borchardt}
	Borchardt, C. W. (1860);\\
	``Über eine Interpolationsformel für eine Art Symmetrischer Functionen und üben Deren Anwendung";\\
	\textit{Math. Abh. der Akademie der Wiessenschaften zu Berlin}, pp. 1--20.
\bibitem{cayley}
	Cayley, A. (1889);\\
	``A theorem on trees";\\
	\textit{Quart. J. Math.}, Vol. 23, pp. 376--8.

\end{thebibliography}
\end{document}